\documentstyle[12pt]{article}
\makeatletter
\@addtoreset{equation}{section}
\makeatother

\newcommand{\AP}{{Ann.\ Phys.\ }}
\newcommand{\APP}{{Acta.\ Phys.\ Pol.\ }}

\newcommand{\EPJ}{{Eur.\ Phys.\ J.\ }}
\newcommand{\JMP}{{J.\ Math.\ Phys.\ }}
\newcommand{\NP}{{Nucl.\ Phys.\ }}
\newcommand{\PL}{{Phys.\ Lett.\ }}
\newcommand{\PR}{{Phys.\ Rev.\ }}
\newcommand{\PRep}{{Phys.\ Rep.\ }}
\newcommand{\PRL}{{Phys.\ Rev.\ Lett.\ }}
\newcommand{\RMP}{{Rev.\ Mod.\ Phys.\ }}
\newcommand{\ZP}{{Z.\ Phys.\ }}
\def\CB {C\llap{/\kern3pt}}
\def\OB {O\llap{/\kern3pt}}
\def\PB {P\llap{/\kern1pt}}
\def\KB {K\llap{/\kern3pt}}
\def\pB {p\llap{/\kern1pt}}
\def\kB {k\llap{/\kern1pt}}
\def\qB {q\llap{/\kern1pt}}
\def\KB {K\llap{/\kern1pt}}
\def\FB {F\llap{/\kern1pt}}
\def\lB {l\llap{/\kern1pt}}
\def\LB {L\llap{/\kern1pt}}
\def\calDB {{\cal D}l\llap{/\kern1pt}}
\def\SigmaB {\Sigma\llap{/\kern1pt}}
\def\OmegaB {\Omega\llap{/\kern1pt}}
\def\barOmegaB {\bar\Omega\llap{/\kern1pt}}
\begin{document}
\thispagestyle{empty}
\vspace*{11mm}
\begin{center}
{\Large \bf Two mechanisms for the elimination of pinch
singularities in out of equilibrium thermal field theories}\\
\vspace*{25mm}
{\large
 I.~Dadi\'c$^{1,2}$ }\footnote{
E-mail: dadic@faust.irb.hr}
\\[24pt]
$^1$ Ruder Bo\v{s}kovi\'{c} Institute,  Zagreb, Croatia\\
$^2$ Fakult\"at f\"ur Physik, Universit\"at Bielefeld , Germany \\
\vspace*{25mm}
\end{center}
\begin{abstract}
We analyze ill-defined pinch singularities characteristic of
out of equilibrium thermal field theories.
We identify two mechanisms that eliminate pinching even at the single
self-energy insertion approximation to the propagator: the first is
based on the vanishing of phase space at the singular point
(threshold effect). It is effective in QED with a massive electron
and a massless photon. In massless QCD, this mechanism fails, but the
pinches cancel owing to the second mechanism, i.e., owing to the
spinor/tensor structure of the single self-energy
insertion contribution to the propagator.
The constraints imposed on distribution functions are very
reasonable.
\end{abstract}
\newpage
\section{ Introduction}
Out of equilibrium thermal field theories have
recently attracted much interest. From the experimental point of
view, various aspects of heavy-ion collisions and the related hot
QCD plasma are of considerable interest, in particular the
supposedly gluon-dominated stage.

Contrary to the equilibrium case$^{(\cite{landsman,mleb})}$ where
pinch, collinear, and infrared problems
have been successfully controlled$^{(\cite{bps,cd,aab,glp})}$,
out of equilibrium
theory$^{(\cite{schwinger,keldysh,rammer})}$ has
suffered from them to these days. However, progress
has been made in this field, too.

Weldon$^{\cite{weldon11}}$ has observed that the out of equilibrium
pinch singularity does not cancel; hence it  spoils analyticity
and causality. The problem gets
worse with more than one self-energy insertions.

Bedaque has argued that in out of equilibrium theory
the time extension should be finite. Thus, the time integration
limits from $-\infty $ to $+\infty $, which are responsible
for the appearance of pinches, have to be abandoned as
unphysical$^{\cite{bedaque}}$.

Le Bellac and Mabilat$^{\cite{lebellac}}$  have
found that collinear singularities cancel in scalar theory,
and in QCD using physical gauges, but not in the case of covariant
gauges.
Ni\'egawa$^{\cite{niegawacom}}$ has found that the pinch-like term
contains a divergent part that cancels collinear singularities
in the covariant gauge.

In their discussion of the pinch-like term
Le Bellac and Mabilat$^{\cite{lebellac}}$ try to avoid the problems with
pinching singularity by substituting the bare retarded
photon (gluon)
propagator with the resummed Schwinger-Dyson series calculated
in the HTL$^{\cite{p,ebp,ft}}$ approximation:
\begin{equation}\label{htlt}
\Delta_T(q_o,q)={-1\over q_o^2-q^2
-m^2\left(x^2+{x(1-x^2)\over 2}\log{x+1\over x-1}\right)},~~
\Delta_T(q_o,q)=
{-1\over q^2+2m^2\left(1-{x\over 2}\log{x+1\over x-1}\right)},
\end{equation}
with the spectral function given by
$\rho_{T,L}=2Im\Delta_{T,L}(q_o+i\eta,q)$.
In the above expressions $m$ is the thermal photon (gluon) mass
given by $m^2=e^2T^2/6$ ($m^2=g^2T^2(N_c+N_f/2)/6$ for the gluon;
note that they assume a
small deviation from equilibrium), and $x=q_o/q$ .
In equilibrium, at high temperatures and low momenta, this substitution
is necessary in order to obtain the results correct to the leading order
in $g^2$. One expects that similar methods work also for a narrow
class of particle distributions corresponding to "high temperatures"
out of equilibrium.
However, their expression for pinch-like contribution differs from the one
following from our general expression for the resummed Schwinger-Dyson
series (\ref{sol1barG}).
At medium and large photon momenta the elimination of pinching by the
use of the HTL approximated propagator is no more justified.

Altherr and Seibert have found that in massive $g^2\phi^3$ theory
pinch singularity does not occur owing to the kinematical
constraint$^{\cite{as}}$.
This result is restricted to the case of
one-loop self-energies.

Altherr has suggested a regularization
method in which the propagator is modified
by the width $\gamma $ which is an arbitrary function of momentum
to be calculated in a self-consistent way. In $g^2\phi^4$ theory,
for small deviations from
equilibrium, $\gamma$ was found to be just the usual equilibrium
damping rate$^{\cite{altherr}}$.

This recipe has been justified in the resummed Schwinger-Dyson
series in various problems with
pinching$^{\cite{bdr,bdrs,bdrk,carrington,niegawa}}$.

Baier, Dirks, and Redlich$^{\cite{bdr}}$ have calculated the
$\pi-\rho $ self-energy contribution to
 the pion propagator,
regulating pinch contributions by the damping rate. In
subsequent papers with
Schiff$^{\cite{bdrs,bdrk}}$ they have calculated the quark propagator
within the HTL approximation$^{\cite{p,ebp,ft}}$; in the resummed
Schwinger-Dyson series, the pinch is naturally regulated by
$Im\Sigma_R$.

Carrington, Defu, and Thoma$^{\cite{carrington}}$ have found that
no pinch singularities appear in the
HTL approximation to the resummed photon propagator .

Ni\'egawa$^{\cite{niegawa}}$ has introduced the notion of renormalized
particle-number density. He has found that, in the appropriately
redefined calculation scheme, the amplitudes and reaction rates are
free from pinch singularities.

By pinching singularity we understand the contour passing
between two infinitely close poles:
\begin{equation}\label{pinch}
\int {dx\over (x+i\epsilon)(x-i\epsilon)},
\end{equation}
where  $x=q^2-m^2 $.
It is controlled by some parameter,
e.g., $\epsilon $. For finite $\epsilon $, the expression is regular.
However, when $\epsilon $ tends to zero, the integration path is
"pinched" between the two poles, and the expression is ill-defined.
Integration gives an $\epsilon^{-1} $ contribution  plus
regular terms. By performing a simple decomposition
of $(x\pm i\epsilon)^{-1}$ into
$PP(1/x)\mp i\pi\delta (x)$, one obtains the related ill-defined
$\delta^2$ expression.

The following expression, which 
is similar to (\ref{pinch}), corresponds
to the resummed Schwinger-Dyson series:
\begin{equation}\label{pinchlikesd}
\int dx{\omega(x)\over (x-\Sigma_R(x)+i\epsilon)
(x-\Sigma^*_R(x)-i\epsilon)},
\end{equation}
where $\omega(x)$ and $\bar \omega(x)$ (which
appears in (\ref{pinchlike1})) are, respectively, proportional to
$\Omega(x)$ and $\bar \Omega(x)$,
where $\Omega(x)$, $\Sigma_R(x)$,  and $\bar \Omega(x)$
are  the components
of the self-energy matrix to be defined in Sec. III.

In expression (\ref{pinchlikesd}), pinching is
absent$^{\cite{bdr,bdrs,bdrk,carrington,niegawa}}$ if
$Im\Sigma_R(x_o)\neq0$
at a value of $x_o$
satisfying $x_o-Re\Sigma_R(x_o)=0$.

The expression corresponding to the single self-energy
insertion approximation to the propagator is similar to
(\ref{pinchlikesd}):
\begin{equation}\label{pinchlike1}
\int dx{\bar \omega(x)\over (x+i\epsilon)(x-i\epsilon)}.
\end{equation}
One can rewrite the integral as
\begin{equation}\label{rpinchlike1}
\int {dx\over 2}\biggl({1\over x+i\epsilon}
+{1\over x-i\epsilon}\biggr)
{\bar \omega(x)\over x}.
\end{equation}
If it happens that
\begin{equation}\label{rpinchlike2}
\lim_{x \rightarrow 0} {\bar \omega(x)\over x}=K<\infty ,
\end{equation}
then  the integral (\ref{rpinchlike1}) decomposes into two pieces
that, although possibly divergent, do not suffer from pinching.

There are two cases in which the function $\bar \omega(x)$ is even 
identically zero in the vicinity of the $x=0$ point:
in thermal equilibrium, because of detailed balance
relations; in massive $g^2 \phi^3$ theory out of equilibrium, 
owing to the mass shell condition$^{\cite{as}}$. The latter mechanism
also works in
out of equilibrium QED if a small photon mass $m_{\gamma} $
is introduced. However, this
elimination of pinching can be misleading: the domain of $x$, where
$\bar \omega(x)=0 $, shrinks to a point as
$m_{\gamma} \rightarrow 0$.
We shall show that the elimination of pinching also occurs
in the $m_{\gamma}=0$ case.

In this paper we identify two mechanisms leading to 
relation (\ref{rpinchlike2}). They are based on the observation that
in the pinch-like contribution loop particles have to be on mass 
shell.

The first mechanism is effective in out of equilibrium QED:
in the pinch-like contribution to the
electron propagator, phase space vanishes linearly
as $x \rightarrow 0$ . In the pinch-like contribution to the photon
propagator, the
domain of integration is shifted to infinity as $x \rightarrow 0$. 
For distributions disappearing rapidly enough at large energies, the
contribution again vanishes linearly in the $x \rightarrow 0$ limit.
This mechanism is also valid in
QCD in the cases with massive quarks.

In out of equilibrium massless QCD, phase space
does not vanish, but
there is an alternative mechanism: the spinor/tensor structure
in all cases leads to relation (\ref{rpinchlike2}).

Also, in out of equilibrium massless QCD, introduction of
a small gluon mass does not help. In this case, processes like
$q\bar q\rightarrow g $ are kinematically allowed, the spinor/tensor
structure is modified, and $\bar \Omega $
does not vanish in the $x \rightarrow 0$ limit.

In a few cases, none of the mentioned mechanisms works and one has
to sum the Schwinger-Dyson series.
This is  the case of the $\pi-\rho $ loop in the
$\pi $ self-energy . Even in the limit of zero pion mass,
$\bar \omega(x)$ vanishes only as $|x|^{1/2}$ and relation 
(\ref{rpinchlike2}) is not fulfilled. A similar problem appears in
electroweak interactions involving
decays of $Z$ and $W$ bosons, decay of Higgs particles, etc.
Another important case is massless $g^2\phi^3$ theory. In contrast to
massless QCD, massless $g^2\phi^3$ theory contains
no spin factor to provide a $q^2$ factor
necessary to obtain (\ref{rpinchlike2}).

The densities are restricted only mildly: they should be cut off
at high energies,
at least as $|k_o|^{-3-\delta}$, in order to obtain a finite total particle
density; for nonzero $k_o$, they should be finite;
for $k_o$ near zero, they should not diverge more rapidly
than $|k_o|^{-1}$, the electron (positron) distribution
should have a finite derivative.
Further restrictions may come from Slavnov-Taylor
identities$^{(\cite
{kkm,kkr,weldont})}$, but they are not crucial for our analysis.

When necessary we assume that the zero-temperature renormalization has
already been performed. The 
"finite temperature"$^{\cite{dh,dhr,keil,bp}}$
 renormalization out of equilibrium$^{\cite{calzetta}}$
may give rise to new problems formally similar to those treated here.
Their treatment is beyond the scope of this paper.

The paper is organized as follows.

In Sec. II we analyze the Schwinger-Dyson equation
in the Keldysh representation, solve it formally, and identify
pinch-like expressions. For one-loop self-energy
insertions, we find that the Keldysh component 
($\bar \Omega (q^2)$) of the self-energy
is responsible for pinches.
We further find that the nonzero Keldysh component requires loop 
particles to be on shell.

In Sec. III we analyze functions such as $\Omega$,
$\bar\Omega$, and $Im\Sigma_R$,
and investigate their threshold properties.

In Sec.IV we show that the electron and photon propagators,
calculated in the single self-energy
insertion approximation, are free from pinching.

In Sec. V we analyze pinch-like expressions in the $q-\bar q$,
$g-g$, and ghost-ghost contributions to the gluon propagator, 
the quark
propagator and the ghost propagator in the single self-energy insertion
approximation. We find that, in all the cases, the spinor/tensor
factor $F$ contains a factor $q^2$ that
is sufficient to eliminate pinching.

In Sec. VI we briefly recollect the main results of the paper.

\section{ Propagators and the Schwinger-Dyson equation}
We start$^{\cite{cshy,niemi}}$ by defining out of equilibrium
thermal propagators for bosons, in the case
when we can ignore the variations of slow variables in Wigner
functions$^{\cite{lebellac,bio}}$:
\begin{equation}\label{D}
D=\left(\matrix{D_{11}&D_{12}\cr
                D_{21}&D_{22}\cr}\right),
\end{equation}
\begin{equation}\label{D11}
D_{11}(k)=D^*_{22}(k)={i \over k^2-m^2+2i\epsilon|k_o|}+
2\pi \sinh^2\theta\delta(k^2-m^2),
\end{equation}
\begin{equation}\label{D12}
D_{12}(k)=2\pi \delta(k^2-m^2)
(\cosh^2 \theta \Theta(k_o)+\sinh^2 \theta \Theta(-k_o)),
\end{equation}
\begin{equation}\label{D21}
D_{21}(k)=2\pi \delta(k^2-m^2)
(\cosh^2 \theta \Theta(-k_o)+\sinh^2 \theta \Theta(k_o)).
\end{equation}

For particles with additional degrees of freedom,  relations
(\ref{D})-(\ref{D21}) are
provided with extra factors $(\kB+m)$ for spin 1/2,
$\left(g_{\mu \nu}-(1-a)k_{\mu}k_{\nu}/(k^2\pm 2i\epsilon k_o)
\right)$ for vector particle,
etc., and similarly for internal degrees of freedom.
To keep the discussion as general as possible,
we show these factors explicitly only when necessary.
The propagator defined by relations (\ref{D})-(\ref{D21}) satisfies
the important condition
\begin{equation}\label{sumD}
0=D_{11}-D_{12}-D_{21}+D_{22}.
\end{equation}

In the case of equilibrium, we have
\begin{equation}\label{eqB}
\sinh^2\theta(k_o)=n_B(k_o)={1 \over \exp\beta |k_o|-1}.
\end{equation}
To obtain the corresponding relations for fermions, we only need
to make the substitution
\begin{equation}\label{b-f}
\sinh^2\theta(k_o) \rightarrow -\sin^2\bar\theta(k_o).
\end{equation}
In the case of equilibrium, for fermions we have
\begin{equation}\label{eqF}
\sin^2\bar\theta_{F,\bar F}(k_o)=
n_{F,\bar F}(k_o)={1 \over \exp\beta (|k_o|\mp\mu)+1}.
\end{equation}
Out of equilibrium, $n_B(k_o)$ and $n_F(k_o)$ will be some given
functions of $k_o$.

To transform into the Keldysh form,
one defines the matrix Q as
\begin{equation}\label{Q}
Q={1 \over \sqrt2}\left(\matrix{-1&1\cr
                               1&1\cr}\right).
\end{equation}
Now
\begin{equation}\label{QDQ}
\left(\matrix{0&D_R\cr
                D_A&D_K\cr}\right)=QDQ^{-1},
\end{equation}
\begin{equation}\label{DR}
D_R(k)=-D_{11}+D_{21}={-i \over k^2-m^2+2i\epsilon k_o},
\end{equation}
\begin{equation}\label{DA}
D_A(k)=-D_{11}+D_{12}={-i \over k^2-m^2-2i\epsilon k_o}
=-D_R^*(k)=D_R(-k),
\end{equation}
\begin{equation}\label{barD}
D_K(k)=D_{11}+D_{22}=2\pi \delta(k^2-m^2)(1+2\sinh^2\theta).
\end{equation}
We need $D_K$ expressed through $D_R$ and $D_A$:
\begin{equation}\label{barDRA}
D_K=h(k_o)(D_R-D_A),~~~
h(k_o)=-\epsilon(k_o)(1+2\sinh^2\theta).
\end{equation}
Again for fermions, $D_K$ is equal to
\begin{equation}\label{barDF}
D_K(k)=D_{11}+D_{22}=2\pi \delta(k^2-m^2)(1-2\sin^2\bar\theta).
\end{equation}
The proper self-energy
\begin{equation}\label{Sigma}
\Sigma=\left(\matrix{\Sigma_{11}&\Sigma_{12}\cr
                \Sigma_{21}&\Sigma_{22}\cr}\right)
\end{equation}
satisfies the condition
\begin{equation}\label{sumSigma}
0=\Sigma_{11}+\Sigma_{12}+\Sigma_{21}+\Sigma_{22}.
\end{equation}
It is also transformed into the Keldysh form (in Niemi's paper there
is a misprint using  $Q^{-1}$ instead of $Q$):
\begin{equation}\label{KSigma}
\left(\matrix{\Omega&\Sigma_A\cr
                \Sigma_R&0\cr}\right)=Q \Sigma Q^{-1},
\end{equation}
\begin{equation}\label{SigmaR}
\Sigma_R=-(\Sigma_{11}+\Sigma_{21}),
\end{equation}
\begin{equation}\label{SigmaA}
\Sigma_A=-(\Sigma_{11}+\Sigma_{12}),
\end{equation}
\begin{equation}\label{Omega}
\Omega=\Sigma_{11}+\Sigma_{22}.
\end{equation}
We also find
\begin{equation}\label{SAR}
\Sigma_A=\Sigma_R^*.
\end{equation}
The "cutting rules" (refs.\cite {weldoncr,ksem}, see also ref.\cite{gelis}
for application of the rules out of equilibrium)
will convince us
that only on-shell loop-particle momenta contribute to
$Im\Sigma_R$ and $\Omega $.

The calculation of the $\Sigma $ matrix gives
(propagators $S(k)$ and $G(k)$ in
the self-energy matrix and in the Schwinger-Dyson equation
are also given by (\ref{D}) to (\ref{barDRA}), with the spin indices
suppressed to keep the discussion as general as possible):
\begin{equation}\label{SigmaAint}
\Sigma_R=-i{1 \over 2}g^2\int {d^4k \over (2\pi)^4}
\big(D_R(k)(S_R(k-q)-S_K(k-q))+(D_A(k)-D_K(k))S_A(k-q)\big),
\end{equation}
\begin{equation}\label{SigmaRint}
\Sigma_A=-i{1 \over 2}g^2\int {d^4k \over (2\pi)^4}
\big((D_R(k)-D_K(k))S_R(k-q)+D_A(S_A(k-q)-S_K(k-q))\big),
\end{equation}
\begin{equation}\label{Omegaint}
\Omega=i{1 \over 2}g^2\int {d^4k \over (2\pi)^4}
\big((D_R(k)+D_A(k))(S_A(k-q)+S_R(k-q))+D_K(k)S_K(k-q)\big).
\end{equation}
%
The Schwinger-Dyson equation
\begin{equation}\label{Schwinger-Dyson}
{\cal G}=G+iG \Sigma {\cal G},
\end{equation}
can be written in the Keldysh form as
\begin{equation}\label{Keldysh}
\left(\matrix{0&{\cal G}_R\cr
                {\cal G}_A&{\cal G}_K\cr}\right)=
\left(\matrix{0&G_R\cr
                G_A&G_K\cr}\right) +
                i\left(\matrix{0&G_R\Sigma_R{\cal G}_R\cr
                G_A\Sigma_A{\cal G}_A&G_A\Omega {\cal G}_R+
                G_K\Sigma_R{\cal G}_R +
                G_A\Sigma_A{\cal G}_K\cr}\right).
\end{equation}

By expanding (\ref{Keldysh}), we deduce the contribution from the
single self-energy insertion to be of the form
\begin{equation}\label{psol2GRA}
{\cal G}_R\approx G_R+iG_R\Sigma_RG_R,~~~
{\cal G}_A\approx G_A+iG_A\Sigma_AG_A,
\end{equation}
which is evidently well defined, and the Keldysh component
suspected for pinching:
\begin{equation}\label{psol2barG}
{\cal G}_K\approx G_K+iG_A\Omega G_R+iG_K\Sigma_RG_R
+iG_A\Sigma_AG_K.
\end{equation}
It is easy to obtain a solution$^{\cite{niemi}}$ for ${\cal G}_R$
and ${\cal G}_A$ using the form (\ref{Keldysh}).
One observes that the equations for ${\cal G}_R$
and ${\cal G}_A$ are simple and the solution is straightforward:
\begin{equation}\label{sol1GR}
{\cal G}_R={1 \over G_R^{-1}-i\Sigma_R}=-{\cal G}_A^*.
\end{equation}
To calculate ${\cal G}_K$, we can use the solution (\ref{sol1GR})
for ${\cal G}_R$ and  ${\cal G}_A$:
\begin{equation}\label{sol1barG0}
{\cal G}_K={\cal G}_A(G_A^{-1}G_KG_R^{-1}+i\Omega) {\cal G}_R.
\end{equation}
Now we eliminate $G_K$ with the
help of (\ref{barDRA}):
\begin{equation}\label{sol1barG}
{\cal G}_K={\cal G}_A\left(h(q_o)(G_A^{-1}-G_R^{-1})
+i\Omega\right){\cal G}_R.
\end{equation}
The first term in (\ref{sol1barG}) is not always zero, but it
does not contain pinching singularities!
The second term in (\ref{sol1barG}) is potentially ill-defined (or
pinch-like).
The pinch-like contribution appears only in this equation;
thus it is the key to the
whole problem of pinch singularities. In the one-loop
approximation, it
requires loop particles to be on mass shell. This will be sufficient
to remove ill-defined expressions in all studied cases.

Equation (\ref{sol1barG}) differs from the one used in
Ref. \cite{lebellac}. Indeed 
in Ref. \cite{lebellac} Im${\cal G}_R$ is used instead of 
${\cal G}_R{\cal G}_A$.

We start with (\ref{psol2barG}).
After substituting (\ref{barDRA})
into (\ref{psol2barG}),
we obtain the regular term plus the pinch-like contribution:
\begin{equation}\label{pert0}
{\cal G}_K\approx {\cal G}_{Kr}+{\cal G}_{Kp},
\end{equation}
\begin{equation}\label{pinch0r}
{\cal G}_{Kr}=h(q_o)\left(G_R-G_A+iG_R\Sigma_RG_R
-iG_A\Sigma_AG_A\right),
\end{equation}
\begin{equation}\label{pinch0p}
{\cal G}_{Kp}= iG_A\bar \Omega G_R,~~~
\bar \Omega=\Omega -h(q_o)(\Sigma_R-\Sigma_A).
\end{equation}
For equilibrium densities, we have
$\Sigma_{21}=e^{-\beta q_o}\Sigma_{12}$ , and
expression (\ref{pinch0p}) vanishes identically.
This is also true for fermions.

Expression (\ref{pinch0p}) is the only one suspected of pinch
singularities at the single self-energy insertion level.
The function $\bar\Omega $ in (\ref{pinch0p}) belongs to the type
of functions characterized by the fact that both loop particles
have to be on mass shell. It is analyzed in detail in Secs. III
and IV (for threshold effect) and in Sec. V (for spin effect).
With the help of this analysis we show that
relation (\ref{pinch0p}) transforms into
\begin{equation}\label{pinch0pd}
{\cal G}_{Kp}= -i{K(q^2,m^2,q_o)\over 2}
\left({1\over q^2-m^2+2i\epsilon q_o}
+{1\over q^2-m^2-2i\epsilon q_o}\right),
\end{equation}
where $K(q^2,m^2,q_o)$ is $\bar\Omega/(q^2-m^2)$ multiplied by
spinor/tensor factors included in the definition of $G_{R,A}$.
The finiteness of the limit
\begin{equation}\label{pinch0pdk}
\lim_{q^2\rightarrow m^2\mp 0}K(q^2,m^2,q_o) = K_{\mp}(q_o) < \infty
\end{equation}
is important for cancellation of pinches. The index $\mp $ indicates
that the limiting value $m^2$ is
approached from either below or above, and these two values  are
generally different.
To isolate the potentially divergent terms, we express the function
$K(q^2,m^2,q_o)$ in terms of functions that are
symmetric ($K_{1}(q^2,m^2,q_o)$)
and antisymmetric ($K_2(q^2,m^2,q_o)$) around the value $q^2=m^2$:
\begin{equation}\label{Kby12}
K(q^2,m^2,q_o)=(K_1(q^2,m^2,q_o)+\epsilon(q^2-m^2)K_2(q^2,m^2,q_o)\big).
\end{equation}
These functions are given by
\begin{equation}\label{K12}
K_{1,2}(q^2,m^2,q_o)={1\over 2}\big(K(q^2,m^2,q_o)\pm K(2m^2-q^2,m^2,q_o)\big).
\end{equation}
Locally (around the value $q^2=m^2$),
these functions are related to the limits $K{\pm}(q_o)$ by
\begin{equation}\label{K12pm}
K_{1,2}(q^2,m^2,q_o)={1\over 2}\big(K_{+}(q_o)\pm K_{-}(q_o)\big).
\end{equation}
As a consequence, the right-hand side of
expression (\ref{pinch0pd}) behaves locally as
\begin{equation}\label{pinchs}
{\cal G}_{Kp}(q^2,m^2,q_o)\approx-{i\over 2}\left(K_1(q_o)+
\epsilon(q^2-m^2)K_2(q_o)\right)
\left({1\over q^2-m^2+2i\epsilon q_o}
+{1\over q^2-m^2-2i\epsilon q_o}\right),
\end{equation}
and the term proportional to $K_2$ is capable of producing
logarithmic singularity.
Furthermore, we were unable to eliminate pinches
related to the double, triple, etc.,
self-energy insertion contributions to the propagator.

\section{Threshold factor}
In this section we analyze the phase space of the loop integral
with both loop particles on mass shell. Special care is devoted
to the behavior of this integral near thresholds. In this analysis
the densities are constrained only  mildly: they are supposed to be
finite and smooth, with a possible exception at zero energy. We  also
assume that the total density of particles is finite.
The expressions are written for all particles being bosons, and
spins are not specified; change to fermions is elementary.

To obtain the integrals over the products of $D_{R,A}$ and
$S_{R,A}$, we start with a useful relation:
\begin{eqnarray}\label{trint0}
& &\int_{-\infty}^{+\infty} dk_o f(k,q){1 \over (k^2-m_D^2+i\lambda k_o\epsilon)
((k-q)^2-m_S^2+i\eta (k_o-q_o)\epsilon)}\cr
\nonumber\\
& &=-i\pi\int_{-\infty}^{+\infty} dk_o f(k,q){\cal P}
\big({\lambda\epsilon(k_o)\delta(k^2-m_D^2)\over
(k-q)^2-m_S^2}
+{\eta\epsilon(k_o-q_o)\delta((k-q)^2-m_S^2)\over
k^2-m_D^2}\big)\cr
\nonumber\\
& &+2\pi^2\delta_{\lambda,-\eta}\int_{-\infty}^{+\infty} dk_o f(k,q)
\epsilon(k_o)\epsilon(k_o-q_o)
\delta(k^2-m_D^2)\delta((k-q)^2-m_S^2)
\end{eqnarray}
where $f(k,q)$, as a function of $k_o$, is some polynomial
of order 0, 1, 2, or 3.

Relation (\ref{trint0}) is obtained as an average of the results
obtained by closing the integration path through the upper
and through the lower semi-plane.

In the defining relations (\ref{D11}) to (\ref{D21})
the particle distribution $\sinh^2\theta $
(and similarly for $\sin^2\bar \theta $)
appears only in the expression where it is multiplied by
$\delta(k^2-m^2)$. Thus there is  freedom to replace $k_o$
by its on mass shell value $\epsilon(k_0)\omega_D$,
where $\omega_D=(\vec k^2-m_D^2)^{1/2}$.
The physical results should not be altered by this replacement.
Then we write $h(k_0)$, defined in (\ref{barDRA}),
as $-k_0/\omega_D(1+2\sinh^2(\omega_D))$,
and similarly for $h(k_o-q_o)$.
This substitution matters when one wants
to perform integrals in (\ref{SigmaAint}) to
(\ref{Omegaint}) with the help of (\ref{trint0}).
With this replacement, $f(k,q)$ in (\ref{trint0}) does not depend on densities
and (\ref{trint0}) can be proved as stated above, without need to discuss the
analytic properties of $\sinh^2\theta $ and $\sin^2\bar\theta $.
Without our replacement, one would be immediately stack with the
question how these quantities (i.e., $\sinh^2\theta $ and
$\sin^2\bar \theta $), supposed to be known, within some uncertainty,
along the real axis,
behave in the complex plane very far from the real axis!

Similar relations could be obtained for higher powers
of $D_{R,A}$ and $S_{R,A}$. For example, for the nth power of
$k^2-m_D^2+i\lambda k_o\epsilon$, the real part of the integral
will be obtained by substituting
$\delta^{(n)}(k^2-m_D^2)(-1)^{(n)}$ instead of
$\delta(k^2-m_D^2)$.
Now we easily calculate $Re\Sigma_R$ as
\begin{equation}\label{trint01}
Re\Sigma_R={-g^2 \over 2(2\pi)^3}\int d^4k
{\cal P}\left({\epsilon(k_o)\delta(k^2-m_D^2)h_D(k_o)
\over ((k-q)^2-m_S^2)}
+{\epsilon(k_o-q_o)\delta((k-q)^2-m_S^2)h_S(k_o-q_o)
\over (k^2-m_D^2)}\right)F.
\end{equation}
$F$ is the factor dependent on spin and
internal degrees of freedom.
The thermal part of $Re\Sigma_R$ is given by:
\begin{equation}\label{trint02}
Re\Sigma_{R~th}={g^2 \over (2\pi)^3}\int d^4k
{\cal P}\left({\delta(k^2-m_D^2)\sinh_D^2(k_o)
\over ((k-q)^2-m_S^2)}
+{\delta((k-q)^2-m_S^2)\sinh_S^2(k_o-q_o)
\over (k^2-m_D^2)}\right)F.
\end{equation}
At equilibrium equation (\ref{trint02}) (after necessary
boson$\rightarrow $fermion conversion) agrees with the known
results$^{\cite{bps,aab}}$.

Now, starting from (\ref{SigmaR}) to (\ref{Omegaint}), we calculate
$\Omega $ and $Im\Sigma_R$.

\begin{equation}\label{Omegai2}
\Omega=2iIm\Sigma_{11}
=2{ig^2 \over 2}\int {d^4k \over (2\pi)^4} 4\pi^2
\delta(k^2-m_D^2)\delta((k-q)^2-m_S^2)N_{\Omega }(k_o,k_o-q_o)F,
\end{equation}
where
\begin{equation}\label{nomega}
N_{\Omega}(k_o,k_o-q_o)=
{1 \over 2}(-\epsilon(k_o(k_o-q_o))+
(1+2\sinh^2\theta_D(k_o))(1+2\sinh^2\theta_S(k_o-q_o))),
\end{equation}
\begin{equation}\label{SigmaRi2}
Im\Sigma_R={g^2 \over 2}\int {d^4k \over (2\pi)^4} 4\pi^2
\delta(k^2-m_D^2)\delta((k-q)^2-m_S^2)
N_R(k_o,k_o-q_o)F,
\end{equation}
and
\begin{eqnarray}\label{nr}
& &N_R(k_o,k_o-q_o)=
(\sinh^2\theta_D(k_o)\epsilon(k_o-q_o)
+\sinh^2\theta_S(k_o-q_o)\epsilon(-k_o)\cr
\nonumber\\
& &+\Theta(-k_o)\Theta(k_o-q_o)-\Theta(k_o)\Theta(q_o-k_o)).
\end{eqnarray}
At equilibrium equations (\ref{nomega}) to (\ref{nr}) agree (after
setting F=-1 for scalar case) with the
corresponding equations obtained for boson-boson
intermediate state$^{\cite{ksem}}$.

It is useful to define $N_{\bar \Omega}(k_o,k_o-q_o)$ as
\begin{equation}\label{nbaromega}
N_{\bar \Omega}(k_o,k_o-q_o)=
N_{\Omega}(k_o,k_o-q_o)-h(q_o)N_R(k_o,k_o-q_o).
\end{equation}
After integrating over $\delta$'s, one obtains expressions of the
general form
\begin{equation}\label{trint1}
{\cal I}={1 \over 4|\vec q|}\int
{|k_o|dk_o \over |\vec k|}d\phi N(k_o,k_o-q_o)
F(q_o,|\vec q|,k_o, |\vec k|, \vec q\vec k, ...)
\Theta(1-z_o^2),
\end{equation}
where $|\vec k|=(k_o^2-m_D^2)^{1/2}$,
\begin{equation}\label{trint2}
\vec q\vec k=|\vec q||\vec k|z_o,
\end{equation}
\begin{equation}\label{trint3}
z_o={\vec q^2+\vec k^2-(\vec q-\vec k)^2 \over 2|\vec k||\vec q|}.
\end{equation}
$\phi\epsilon (0,2\pi)$ is the
angle between vector $\vec k_T$ and $x$ axes.

Let us start with the $q^2>0$ case.
Solution of $\Theta(1-z_o^2)$ gives the integration limits
\begin{equation}\label{ko12}
k_{o~1,2}={1 \over 2q^2}\left(q_o(q^2+m_D^2-m_S^2)\mp
|\vec q|((q^2-q^2_{+tr})(q^2-q^2_{-tr}))^{1/2}\right),
\end{equation}
or
\begin{equation}\label{k12}
|\vec k|_{1,2}={1 \over 2q^2}\left(|\vec q|(q^2+m_D^2-m_S^2)\mp
q_o((q^2-q^2_{+tr})(q^2-q^2_{-tr}))^{1/2}\right),
\end{equation}
\begin{equation}\label{qmp}
q_{\pm tr}=|m_D \pm m_S|.
\end{equation}
Assume now that $q_{tr}\neq 0$. In this case, at the
threshold, the limits shrink to the value
\begin{equation}\label{koko}
k_{o~tr}={q_o(q^2_{tr}+m_D^2-m_S^2) \over 2q^2_{tr}},~~~
|\vec k|_{tr}={|\vec q|(q^2_{tr}+m_D^2-m_S^2) \over 2q^2_{tr}}.
\end{equation}
Near the threshold, it is convenient to replace
 the integration variable
by $dk_o|k_o|/|\vec k|=d|\vec k|$. Now, for $|q^2-q^2_{tr}|$ such small,
that the integration limits $k_{o~1,2}$ 
are both of the same sign as $k_{o tr}$,
we have $k_o=\epsilon(k_{o,tr})(\vec k^2+m_D^2)^{1/2}$.

We define the coefficient $c_1$ by
\begin{equation}\label{c1}
c_1={1 \over 4|\vec q|}\int d\phi N(k_{o~tr},k_{o~tr}-q_{o})
F(q_o,|\vec q|,k_{o~{tr}}, |\vec k|_{tr}, \vec q\vec k_{tr}, ...).
\end{equation}
Now the expression (\ref{trint1}) can be approximated by
\begin{eqnarray}\label{2}
& &{\cal I}\approx c_1(|\vec k|_{2}-|\vec k|_{1})\cr
\nonumber\\
& &\approx c_1(\Theta(q^2-q_{+tr}^2)+ \Theta(-q^2+q_{-tr}^2))
{q_o((q^2-q^2_{+tr})(q^2-q^2_{-tr}))^{1/2} \over q^2}.
\end{eqnarray}
Relation (\ref{2}) is the key to further discussion of the
threshold effect.

We obtain this also for higher dimension (D=6, for example).

Relation (\ref{2}) put some limits on the behavior of
density functions:
they should not tend to infinity at any value of $q_o\neq 0$; near
$q_o=0$, owing to the presence of the factor $q_o$,
they should not rise more rapidly than $q_o^{-1}$.

Owing to (\ref{qmp}) and (\ref{2}), the function
${\cal I}(q^2,m_D^2,m_S^2)$
has the following properties
important for cancellation of pinches.

It vanishes between the thresholds, i.e., the domain
$(m_D-m_S)^2<q^2<(m_D+m_S)^2$ is forbidden (${\cal I}=0$). If it happens
that the bare mass $m^2$ belongs to this domain, the single self-energy
insertion will be free of pinching. In this case, multiple
(double, triple, etc.) self-energy insertions will
 also be free of pinching. Massive $\lambda\phi^3$ theory$^{\cite{as}}$
 is a good example of this case.

It is (in principle) different from zero in the allowed domain
$q^2<(m_D-m_S)^2$ and $(m_D+m_S)^2<q^2$. In this case, one cannot get rid
of pinching.
 This situation appears in the $\pi-\rho $
interaction$^{\cite{bdr}}$. An exception to this rule
are occasional zeros owing to the
specific form of densities.

The behavior at the boundaries (i.e., in the allowed region near
the threshold) depends on the masses $m_D$ and $m_S$ and there are
a few possibilities.

If both masses are nonzero and different
($0\neq m_D\neq m_S\neq 0$), then there are two
thresholds and ${\cal I}$ behaves as $(q^2-q_{\pm tr}^2)^{1/2}$
in the allowed region near the threshold $q_{\pm tr}^2$. For
$m^2=q_{tr}^2$, the power
$1/2$ is not large enough to suppress pinching.

If one of the masses is zero ($m_D\neq0, m_S=0$ or
$m_D=0, m_S\neq 0$), then (\ref{2}) gives that the thresholds are
identical (i.e., the forbidden domain shrinks to zero) and one obtains
the $(q^2-m_D^2)^1$ behavior near $m_D^2$. This case (for $m^2=m_D^2$)
is promising. The elimination of pinching in the electron propagator,
considered in Sec.IV, is one of important examples.

If the masses are equal but different from zero
($m_D=m_S\neq 0$), then there are two
thresholds with different behavior. The function ${\cal I}$
behaves as $(q^2-q_{+tr}^2)^{1/2}$
in the allowed region near the threshold $q_{+tr}^2=4m_D^2$,and this
behavior cannot eliminate pinching in the supposed case $m^2=4m_D^2$ .

However, at the other threshold, namely at
 $q^2_{-tr}=0$, the physical region is determined
by $q^2<0$ and the above discussion does not apply.
In fact, the integration limits (\ref{ko12}) or (\ref{k12}) are
valid, but the region between $k_{o~1}$ and $k_{o~2}$
is now excluded from integration.
One has to integrate  over the domain
$(-\infty,k_{o~1})\bigcup (k_{o~2},+\infty)$.
This leads to the limitation in the high-energy behavior of the
density functions. An important example of such behavior, elimination of
pinching in the photon propagator ($m_{\gamma}$), is discussed in Sec.IV.

If both masses vanish ($m_D=m_S=0$), the thresholds coincide,
there is no forbidden region and no threshold behavior. The
behavior depends on the spin of the particles involved.
For
scalars, the leading term in the expansion of ${\cal I}$ does
not vanish. Pinching is not eliminated.

The case of vanishing masses ($m_D=m_S=0$) for particles with spin
exhibits a peculiar behavior. In all studied examples
(see Sec.V for details),
 ${\cal I}$ behaves as $q^2$ as $q^2\rightarrow 0$, which promises
the elimination of pinching.

\section{ Pinch Singularities in QED}
\subsection{ Pinch Singularities in the Electron Propagator}
In  this subsection we apply the results of preceding section
to cancel the pinching singularity appearing in a single
self-energy insertion approximation to the electron propagator.
To do so, we have to substitute $m_D=m$, $m_S=0$,
$\sinh^2\theta_D(k_o)\rightarrow -n_e(k_o)$,
$\sinh^2\theta_S(k_o-q_o) \rightarrow n_\gamma(k_o-q_o)$,
and $h(k_o)=-\epsilon(k_o)(1-2n_e(k_o))$,
where $n_e$ and $n_\gamma$ are given non-equilibrium distributions of
electrons and photons in relations (\ref{nomega}),
(\ref{nr}), (\ref{nbaromega}), and (\ref{barDRA}).
The thresholds are now identical
\begin{equation}\label{qemp}
q^2_{\pm tr}=m^2,
\end{equation}
and the integration limits are
\begin{equation}\label{keo12}
k_{o~1,2}={1 \over 2q^2}\left(q_o(q^2+m^2)\mp
|\vec q|((q^2-m^2)\right)
\end{equation}
or
\begin{equation}\label{ke12}
|\vec k|_{2}-|\vec k|_1={q_o \over q^2}(q^2-m^2)).
\end{equation}
At threshold the limits shrink to the value
\begin{equation}\label{kolim}
k_{o~tr}=q_o,~~~~
|\vec k|_{tr}=|\vec q|.
\end{equation}

Then, with the help of (\ref{c1}), we define
\begin{eqnarray}\label{KB}
& &\KB(q^2,m^2,q_o)={(\qB+m)\barOmegaB(\qB+m)\over (q^2-m^2)}\approx
{1 \over 16\pi^2|\vec q|(q^2-m^2)}\int d\phi N_{\bar \Omega}
(k_{otr}=q_o,k_{otr}-q_o=0)
\cr
\nonumber\\
& &
(\qB+m)\FB(q_o,|\vec q|,k_{otr}, |\vec k|_{tr},
\vec q\vec k_{tr}, ...)(\qB+m)
(|\vec k|_{2}-|\vec k|_{1}).
\end{eqnarray}
The trace factor $\FB $ is calculated with loop particles on mass
shell:
\begin{eqnarray}\label{ftrel}
\FB_{e\gamma}&=&
\left (g_{\mu \nu}-(1-a){(k-q)_{\mu}(k-q)_{\nu}
\over (k-q)^2\pm 2i(k_o-q_o)\epsilon}\right )
\gamma^\mu (\kB+m)\gamma^\nu\cr
\nonumber\\
&=&
\left(-2\kB+4m-
(1-a)(-\qB+m-{(\kB-\qB)(-k^2)q^2 \over (k-q)^2\pm 2i(k_o-q_o)
\epsilon}\right).
\end{eqnarray}
In calculating the term proportional to $(1-a)$, we have to use 
the trick
\begin{equation}\label{mgamma}
((k-q)^2\pm i\epsilon)^{-2}=
\lim_{m_\gamma \rightarrow 0}\left[
{\partial \over \partial m_\gamma^2}
((k-q)^2\pm i\epsilon)-m_\gamma^2)^{-1}\right].
\end{equation}
For $q^2\neq 0$, we can decompose the vector $k$ as
\begin{equation}\label{kl}
k={(k.q)\over q^2}q+{(k.\tilde q)\over \tilde q^2}\tilde q+k_T
=(q-{q_o\over |\vec q|}\tilde q){-m_\gamma^2+m^2+q^2\over 2q^2}+
{k_o\over |\vec q|}\tilde q+k_T,
\end{equation}

where, in the heat-bath frame with the $z$ axis oriented along
the vector $\vec q$, we have
\begin{equation}\label{tildeq}
q=(q_o,0,0,|\vec q|),~~
\tilde q=(|\vec q|, 0,0,q_o),~~
q\tilde q=0,~~\tilde q^2=-q^2,
\end{equation}
The transverse component of $k$, $k_T$ vanishes after
integration over $\phi $.

Finally, we obtain
\begin{eqnarray}\label{qfq}
& &(\qB+m)\FB(\qB+m)=2m(q^2+m^2+2m\qB)\cr
\nonumber\\
& &
+(q^2-m^2)\biggl(-{q^2-m^2\over q^2}\qB
+(-{q_o(q^2+m^2)\over q^2|\vec q|}+2{k_o\over |\vec q|})
\tilde \qB+2\kB_T
\cr
\nonumber\\
& &-(1-a){(q^2-m^2)\over 2q^2}(-\qB+{q_o\over |\vec q|}
\tilde \qB)\biggr).
\end{eqnarray}
Now we can study the limit
\begin{eqnarray}\label{KkB}
\KB(q_o)&=&\lim_{q^2\rightarrow m^2}\KB(q^2,m^2,q_o)
\cr
\nonumber\\
&=&
(\qB+m){q_o \over 2\pi|\vec q|m^2}N_{\bar \Omega}
(k_{o~tr},k_{o~tr}-q_{o}).
\end{eqnarray}
It is easy to find that $\KB(q_o) $ is finite provided that 
$m^2\neq 0$ and
$N_{\bar \Omega}(q_o,0)<\infty $. The last condition
is easy to investigate using the limiting procedure:
\begin{eqnarray}\label{egamma}
& &N_{\bar \Omega}(q_o,0)=\lim_{k_o\rightarrow q_o}
N_{\bar \Omega}(k_o,k_o-q_o)=
\lim_{k_o\rightarrow q_o}2n_\gamma(k_o-q_o)(n_e(q_o)-n_e(k_o))
\cr
\nonumber\\
& &
+\lim_{k_o\rightarrow q_o}\left((n_e(q_o)-n_e(k_o)
-(\epsilon(q_o)\epsilon(k_o-q_o)(n_e(q_o)+n_e(k_o)-2n_e(q_o)n_e(k_o))
\right).
\end{eqnarray}
One should observe here that the integration limits imply that
the limit $k_o\rightarrow q_o$ is taken from below for $q^2>m^2$, 
and from above for $q^2<m^2$. The two limits lead to different
values of $N_{\bar \Omega}(q_o,0)$.
Only the first term in (\ref{egamma}) can give rise to problems. We 
rewrite it as $\lim_{k_o\rightarrow 0}\left(2k_on_\gamma(k_o)
{\partial n_e(k_o+q_o)\over \partial k_o}\right)$.
As relation
(\ref{KkB}) should be valid at any $q_o$, we can integrate over $q_o$
to find that the photon distribution should not grow
more rapidly than
$|k_o|^{-1}$ as $k_o$ approaches zero,
while the derivative of the electron distribution $n_e(q_o)$ should
be finite at any $q_o$:
\begin{equation}\label{cgamma}
\lim_{k_o\rightarrow 0}k_on_\gamma(k_o)<\infty,
\end{equation}
\begin{equation}\label{cel}
|{\partial n_e(q_o)\over \partial q_o}|<\infty.
\end{equation}
Under the very reasonable conditions (\ref{cgamma}) and (\ref{cel})
the electron propagator is free from pinches.

It is interesting to observe the discontinuity of
$\KB(q^2,m^2,q_o)$ at the point $q^2=m^2$. This feature will be
repeated in massless QCD.

It is worth observing that $\KB(q_o) $ is gauge independent, at 
least within the class of covariant gauges.

\subsection{ Pinch Singularities in the Photon Propagator}
To consider the pinching singularity appearing in a single
self-energy insertion
approximation to the photon propagator, we have to make
the substitutions
$m_D=m=m_S$, $\sinh^2\theta_D(k_o)\rightarrow -n_e(k_o)$,
$\sinh^2\theta_S(k_o-q_o) \rightarrow -n_e(k_o-q_o)$,
and $h(k_o)=-\epsilon(k_o)(1+2n_\gamma(k_o))$.
There are two thresholds, but only $q^2_{1, tr}=0$ and the domain 
where $q^2<0$ are relevant  to a
massless photon.
The integration limits are given by the same expression 
(\ref{keo12}), but now we have to integrate over 
the domain $(-\infty,k_{o~1})\bigcup (k_{o~2},+\infty)$.
As $q^2 \rightarrow -0$, we
 find $(k_{o~1}\rightarrow -\infty)$ and
$(k_{o~2}\rightarrow +\infty)$.
The integration domain is still infinite but is shifted toward
$\pm \infty$ where one expects that the particle 
distribution vanishes:
\begin{eqnarray}\label{Kmn}
& &K_{\mu\nu}(q^2,q^o)=
\left(g_{\mu\rho}-(1-a)
{q_{\mu}q_{\rho}\over q^2-2iq_o\epsilon}\right )
{\bar\Omega^{\rho\sigma}\over q^2}
\left(g_{\sigma \nu}-(1-a)
{q_{\sigma}q_{\nu}\over q^2+2iq_o\epsilon}\right )
\cr
\nonumber\\
& &=
{1 \over 16\pi^2|\vec q|q^2}
\left(\int_{-\infty}^{k_{o1}}+\int_{k_{o2}}^{\infty} 
\right){k_odk_o\over |\vec k|}
\int d\phi N_{\bar \Omega}(k_o,k_o-q_o)
\left(g_{\mu\rho}-(1-a)
{q_{\mu}q_{\rho}\over q^2-2iq_o\epsilon}\right )
\cr
\nonumber\\
& &
F^{\rho\sigma}(q_o,|\vec q|,k_o,|\vec k|,\vec q\vec k, ...)
\left(g_{\sigma \nu}-(1-a)
{q_{\sigma}q_{\nu}\over q^2+2iq_o\epsilon}\right ).
\end{eqnarray}
To calculate $F^{\mu\nu}$ for the $e-\bar e$ loop, we
parametrize the loop momentum $k$ by introducing
an intermediary variable $l$ perpendicular to $q$.
$m$ is the mass of loop particles:
\begin{equation}\label{lm}
k=\alpha q+l,~q.l=o,~
k^2=(k-q)^2=m^2,~
l^2=m^2-\alpha^2q^2,~\alpha={k^2+q^2-(k-q)^2 \over 2q^2}.
\end{equation}
At the end of the calculation we eliminate $l$ in favor of $k$.
After all possible singular denominators are canceled,
one can set $\alpha=1/2$.
\begin{eqnarray}\label{fmn}
& &F_{e\bar e}^{\mu \nu}=
-Tr(\kB +m)\gamma^{\mu}(\kB-\qB+m)\gamma^{\nu}\cr
\nonumber\\
& &
=(2q^2g^{\mu \nu}-
2q^{\mu}q^{\nu}+8l^{\mu}l^{\nu})=
\biggl({4m^2q_o^2\over \vec q^2}A^{\mu \nu}(q)\cr
\nonumber\\
& &
+{q^2 \over \vec q^2}\biggl((4k_o(k_o-q_o)-4m^2-q^2)A^{\mu \nu}(q)
+(-8(k_o-{q_o\over 2})^2+2\vec q^2) B^{\mu \nu}(q)\biggr)\biggr).
\end{eqnarray}

Using relation (\ref{gfg}) we obtain
\begin{eqnarray}\label{Kmns}
& &K_{\mu\nu}(q^2,q_o)=
{1 \over 16\pi^2|\vec q|q^2}
\left(\int_{-\infty}^{k_{o1}}+\int_{k_{o2}}^{\infty} \right)
{k_odk_o\over |\vec k|}
\int d\phi N_{\bar \Omega}(k_o,k_o-q_o)
\cr
\nonumber\\
& &
\biggl({4m^2q_o^2\over \vec q^2}A_{\mu \nu}(q)\cr
\nonumber\\
& &
+{q^2 \over \vec q^2}\biggl((4k_o(k_o-q_o)-4m^2-q^2)A_{\mu \nu}(q)
+(-8(k_o-{q_o\over 2})^2+2\vec q^2) B_{\mu \nu}(q)\biggr)\biggr).
\end{eqnarray}
In the integration over $k_o$ the terms proportional to
$(k_o^2q^2)^n$ dominate and 
$\lim_{q^2\rightarrow 0}|K_{\mu\nu}(q^2,q_o)|<\infty$
if
\begin{eqnarray}\label{Kmna}
|{1 \over 16\pi^2|\vec q|q^2}
\left(\int_{-\infty}^{k_{o1}}+\int_{k_{o2}}^{\infty} \right)
{k_odk_o\over |\vec k|}(\alpha+\beta k_o^2 q^2)
\int d\phi N_{\bar \Omega}(k_o,k_o-q_o)|<\infty.
\end{eqnarray}
Here $N_{\bar \Omega}(k_o,k_o-q_o)$ is given by
\begin{eqnarray}\label{ebare}
& &N_{\bar \Omega}(k_o,k_o-q_o)=
-2n_{e}(k_o-q_o)(-n_\gamma(q_o)-n_e(k_o))
\cr
\nonumber\\
& &
-n_\gamma(q_o)-n_e(k_o)
-\epsilon(q_o)\epsilon(k_o-q_o)(-n_\gamma(q_o)
+n_e(k_o)+2n_\gamma(q_o)n_e(k_o)).
\end{eqnarray}
Assuming that the distributions obey the inverse-power law at large 
energies $n_{e}(k_o)\propto |k_o|^{-\delta_e}$ and
$n_{\bar e}(k_o)\propto |k_o|^{-\delta_{\bar e}}$, we find that
the terms linear in densities dominate. Thus, for $n=0,1$, one finds
\begin{equation}\label{ltsa}
{-1\over q^2}\biggl(\int_{-\infty}^{k_{o~1}}
+\int_{k_{o~2}}^{+\infty}\biggr)
{|k_o|dk_o\over |\vec k|}|k_o|^{2n-\delta} (-q^2)^n
\propto (\delta-1-2n)^{-1}(|\vec q|m)^{1+2n-\delta}
(-q^2)^{(\delta-3)/2}.
\end{equation}
It follows that (\ref{Kmna}) is finite (in fact, it vanishes)
if $\delta_e,\delta_{\bar e}>3$. Similar analysis for electron
propagator at $q^2 < 0$ (thus outside of our analysis of
pinch singularities) leads to $\delta_{\gamma}>3$. 
This is exactly the condition
\begin{equation}\label{fine}
\int d^3kn_{\gamma,e,\bar e}(k_o)<\infty.
\end{equation}
Thus the pinching singularity is canceled in the photon propagator
under the condition that the electron and positron distributions
should be such that the total number of particles is finite.

Also, in the photon propagator,
the quantity $\lim_{q^2 \rightarrow 0}K_{\mu\nu}(q^2,q_o)$
does not depend on the gauge parameter.

Expression (\ref{ltsa}) is not valid for $m=0$.

\section{ Pinch Singularities in Massless QCD}
In this section we consider the case of massless QCD.
Pinching singularities, related to massive quarks,
are eliminated by the methods used in the preceding section.

In self-energy insertions related to
gluon, quark, and ghost propagators, the masses in the loop as well
as the masses of the propagated particles are zero.
Thus, the methods of the preceding section do not produce
the expected result. Attention is turned to
the spin degrees of freedom, i.e., to the function $F$ of the 
integrand in (\ref{Omegai2}) to (\ref{trint1}). In the calculation 
of $F$ it has been anticipated that the loop particles have to be
on mass shell. In this case, $F$ provides an extra $q^2$ factor
in all the cases considered, in which not all particles are scalars.
This $q^2 $ factor suffices for the elimination of pinching
singularities.

The integration limits are now
\begin{equation}\label{ko12qg}
k_{o~1,2}={1\over 2}\left(q_o\mp|\vec q|\right).
\end{equation}
The difference $|\vec k|_2-|\vec k|_1$ is finite and
there is no threshold effect.

It is worth observing that for $q^2>0$, we have to integrate between
$k_{o1}$ and $k_{o2}$, whereas for $q^2<0$, the integration domain is
$(-\infty,k_{o~1})\bigcup (k_{o~2},+\infty)$. This leads to two
limits, $\lim_{q^2\rightarrow \pm0}K(q^2,q_o)=K_{\pm}(q_o)$,
in all cases of massless QCD.

By inspection of the final results (\ref{qqse}),(\ref{ghghse}), and
(\ref{ggse}),  we find that the case $q^2<0$ requires integrability 
of the function $k_o^2N_{\bar\Omega}(k_o,k_o-q_o)$ leading to the
condition (\ref{fine})  on the quark, gluon, and ghost distribution
functions.

By using (\ref{lm}), we again introduce the intermediary variable
$l$ perpendicular to $q$; now we have to set $m=0$.

\subsection{ Self-Energy Insertions Contributing to the Gluon
Propagator}
The function $K_{\mu \nu}(q^2,q_o)$ related to the gluon propagator
is the sum
\begin{eqnarray}\label{Kgsum}
& &K_{\mu \nu}(q^2,q_o)=
\Sigma_iK_{q_i\bar q_i massive,\mu \nu}(q^2,q_o)
\cr
\nonumber\\
& &+\Sigma_iK_{q_i\bar q_i massless,\mu \nu}(q^2,q_o)
+K_{ghgh,\mu \nu}(q^2,q_o)+K_{gg,\mu \nu}(q^2,q_o),
\end{eqnarray}
where the terms in the sum are defined as
\begin{eqnarray}\label{VCbarOS}
K_{\mu \nu}(q^2,q_o)=
(g_{\mu\rho}-(1-a)D_{R\mu\rho}(q))
{\bar\Omega^{\rho \sigma} \over q^2}
(g_{\sigma\nu}-(1-a)D_{A\sigma\nu}(q)).
\end{eqnarray}
Pinching singularities, related to massive quarks,
are eliminated by the methods used in the preceding section.
The tensor $F$ related to the massless quark-antiquark
contribution to the gluon self-energy is
\begin{eqnarray}\label{qqse}
& &F_{q \bar q}^{\mu \nu}=
-{\delta_{a b} \over 6}Tr\kB\gamma^{\mu}(\kB-\qB)\gamma^{\nu}
={\delta_{a b}\over 6}(2q^2g^{\mu \nu}-
2q^{\mu}q^{\nu}+8l^{\mu}l^{\nu})
\cr
\nonumber\\
& &={\delta_{a b} \over 6}
\biggl({q^2 \over \vec q^2}\biggl((4k_o(k_o-q_o)-q^2)A^{\mu \nu}(q)
+(-8(k_o-{q_o\over 2})^2+2\vec q^2) B^{\mu \nu}(q)\biggr)
+O^{\mu\nu}(\vec k_T)\biggr).
\end{eqnarray}
As $F_{\mu\nu}$ contains only $A$ and $B$ projectors, relation
(\ref{gfg})
guarantees that the result does not depend on the gauge parameter.

Relation (\ref{qqse}) contains only terms proportional to $q^2$,
and $\lim_{q^2\rightarrow 0}K_{\mu\nu}(q^2,q_o)$ is finite.

For the ghost-ghost contribution to the gluon self-energy, 
the tensor $F$ is given by
\begin{eqnarray}\label{ghghse}
& &F_{gh gh}^{\mu \nu}=-\delta_{a b}N_ck^{\mu}(k-q)^{\nu}
=-\delta_{a b}N_c\left(-{q^{\mu}q^{\nu} \over 4}+l^{\mu}l^{\nu}
+{q^{\mu}l^{\nu}-l^{\mu}q^{\nu} \over 2}\right)\cr
\nonumber\\
& &=
-\delta_{a b}N_c{q^2 \over \vec q^2}
\biggl({4k_o(k_o-q_o)+q^2 \over 8}A^{\mu \nu}(q)
-(k_o-{q_o \over 2})^2 B^{\mu \nu}(q)
-{\vec q^2 \over 4}D^{\mu \nu}(q)+O^{\mu\nu}(\vec k_T)
 \biggr)
.
\end{eqnarray}
The antisymmetric part vanishes after integration, so we have
left it out from
the final result in (\ref{ghghse}).

The tensor $F$ for the gluon-gluon contribution to the gluon 
self-energy is
\begin{eqnarray}\label{ggse}
& &F_{gg}^{\mu \nu}=
{\delta_{ab}N_c \over 2}\left(g^{\mu \sigma}(q+k)^{\tau}-
g^{\sigma \tau}(2k-q)^{\mu}+
g^{\tau \mu}(k-2q)^{\sigma}\right)\cr
\nonumber\\
& &
\left(g_{\sigma \sigma'}-(1-a){(k-q)_{\sigma}(k-q)_{\sigma '}
\over (k-q)^2\pm 2i(k_o-q_o)\epsilon}\right)\cr
\nonumber\\
& &\left(g^{\nu \sigma '}(q+k)^{\tau'}-
g^{\sigma '\tau '}(2k-q)^{\nu}+
g^{\tau '\nu}(k-2q)^{\sigma '}\right)
\left(g_{\tau \tau '}-(1-a){k_{\tau}k_{\tau '}
\over k^2\pm 2ik_o\epsilon}\right)\cr
\nonumber\\
& &={\delta_{ab}N_c \over 2}\biggl(
4q^2g_{\mu \nu}- {9 \over 2}q_{\mu}q_{\nu}+10l_{\mu}l_{\nu}
-(1-a)(-5q_\mu q_\nu+3q^2g_{\mu\nu})
\cr\nonumber\\& &
-{1-a\over k^2\pm 2ik_o\epsilon}\biggl(-{q^2\over 4}q_\mu q_\nu
+5q^2l_\mu l_\nu+q^2(l_\mu q_\nu+l_\nu q_\mu)+{q^4 \over 4}
g_{\mu\nu}\biggr )
\cr\nonumber\\& &
-{1-a\over (k-q)^2\pm 2i(k_o-q_o)\epsilon}
\biggl(-{q^2\over 4}q_\mu q_\nu+5q^2l_\mu l_\nu
-q^2(l_\mu q_\nu+q_\mu l_\nu)+{q^4\over 4}g_{\mu\nu} \biggr)
\cr
\nonumber\\
& &
+(1-a)^2\biggl(-2q_\mu q_\nu
+({(1-a)^2\over k^2\pm 2ik_o\epsilon)}-
{(1-a)^2\over (k-q)^2\pm 2i(k_o-q_o)\epsilon})
2q^2(q_\mu l_\nu+l_\mu q_\nu)
\cr
\nonumber\\
& &
+{(1-a)^2\over (k^2\pm 2ik_o\epsilon)((k-q)^2\pm 2i(k_o-q_o)
\epsilon)}
4q^4l_\mu l_\nu
\biggr)
\cr
\nonumber\\
& &
\rightarrow{\delta_{ab}N_c q^2\over 2}\biggl({1\over \vec q^2}
\biggl((10(k_o-{q_o \over 2})^2+{3 \over 2}\vec q^2 )A^{\mu \nu}(q)
\cr
\nonumber\\
& &
+(-10(k_o-{q_o\over 2})^2+4\vec q^2) B^{\mu \nu}(q)
-{\vec q^2 \over 2}D^{\mu \nu}(q)\biggr)
\cr
\nonumber\\
& &
-(1-a)\biggl({1\over 2}A^{\mu\nu}-B^{\mu\nu}
-{q_o\over |\vec q|}C^{\mu\nu} \biggr)
\cr
\nonumber\\
& &
+(1-a)^2\biggl(-{q^2\over \vec q^2}A^{\mu\nu}
+2{q_o^2\over \vec q^2}B^{\mu\nu}
-2{q_o\over |\vec q|}C^{\mu\nu}-2D^{\mu\nu}\biggr)
+O^{\mu\nu}(\vec k_T)\biggr).
\end{eqnarray}
Expressions (\ref{qqse}), (\ref{ghghse}), and (\ref{ggse})
 have been obtained by
substitution of (\ref{llAB}), (\ref{qqDB}), and
(\ref{gAB}) and, finally, by eliminating the
variable $l$ in favor of $k$. The tensor $O^{\mu\nu}(\vec k_T)$ is
linear in $\vec k$, thus it vanishes after integration over $\phi $.

We note here that, in the Feynman gauge ($a=1$), the operator $C$
is absent from the gluon self-energy! Consequently, the relation
originating from Slavnov-Taylor identities (proved in
\cite{kkm,kkr,weldont} for equilibrium
densities), $\Pi^2_C=(q^2-\Pi_L)\Pi_D$, is fulfilled at $a=0$ only if
$\Pi_D=0$. Thus the contributions to $\pi_D$ from the ghost-ghost 
and gluon-gluon self-energies mutually cancel, imposing restrictions
on the densities related to unphysical degrees of freedom. As it
does not interfere with the cancellation of pinches, the problem of
unphysical degrees of freedom will be discussed
elsewhere.

Finally, we need (\ref{gfg}) in all three cases.

Expressions (\ref{qqse}), (\ref{ghghse}), and (\ref{ggse})
for the ghost-ghost, quark-antiquark, and gluon-gluon
contributions to the gluon self-energy contain only terms
proportional to $q^2$. The function $K_{\mu\nu}(q^2,q_o)$
approaches the finite value $K_{\mu\nu}(\pm,q_o)$.

Thus we have shown that the single
self-energy contribution to the gluon propagator is free from pinching
under the condition (\ref{fine}) .

\subsection{ Quark-Gluon Self-Energy
Contribution to the Quark Propagator}
The $K$ spinor for the quark-gluon contribution
to the massless quark propagator is defined as
\begin{equation}\label{SKbarOS}
\KB(q^2,q_o)=\qB{\barOmegaB\over q^2}\qB.
\end{equation}
In the self-energy of a massless quark coupled to a
gluon the spin factor $F$ is given by
\begin{eqnarray}\label{ftrace}
\FB_{qg}&=&\delta_{a b}{N_c^2-1 \over 2N_c}
\left (g_{\mu \nu}-(1-a){(k-q)_{\mu}(k-q)_{\nu}
\over (k-q)^2\pm 2i(k_o-q_o)\epsilon}\right )
\gamma^\mu \kB\gamma^\nu\cr
\nonumber\\
&=&\delta_{a b}{N_c^2-1 \over 2N_c}
\left(-2\kB-
(1-a)(-\qB-{(\kB-\qB)(q^2-k^2) \over (k-q)^2\pm 2i(k_o-q_o)
\epsilon}\right)
\cr
\nonumber\\
&\rightarrow &
\delta_{a b}{N_c^2-1 \over 2N_c}
\left(-\qB+{q_o\over |\vec q|}\tilde\qB
-2{k_o\over |\vec q|}\tilde\qB-2\kB_T
-{1-a\over 2}(-\qB-{q_o\over |\vec q|}\tilde\qB)\right).
\end{eqnarray}
For our further discussion, we need the product

\begin{equation}\label{qtrl}
\qB \FB_{qg}\qB =\delta_{a b}{N_c^2-1 \over 2N_c}q^2
\left(-\qB-{q_o\over |\vec q|}\tilde\qB+2{k_o\over |\vec q|}
\tilde\qB+2\kB_T
-{1-a\over 2}(-\qB+{q_o\over |\vec q|}\tilde\qB)\right),
\end{equation}
which contains the damping factor $q^2$.
The term $\kB_T$ will be integrated out.

By inserting (\ref{qtrl}) into (\ref{SKbarOS}), we
obtain (\ref{pinch0pdk}) free from pinches.

To calculate $\KB(q_o)$,
we need the limit
\begin{equation}\label{qtrll}
\lim_{q^2\rightarrow 0}{\qB \FB_{qg}\qB\over q^2}
=\delta_{a b}{N_c^2-1 \over 2N_c}{2(k_o-q_o)\over q_o}\qB.
\end{equation}
From (\ref{qtrll}) we conclude that $\KB(q_o)$ does not depend
on the gauge parameter.

Omitting details, we observe that pinching is absent from
the quark propagator, also in the Coulomb gauge.

\subsection{ Ghost-Gluon Self-Energy
Contribution to the Ghost Propagator}
The $K$ factor is defined as
\begin{equation}\label{SgKbarOS}
K(q^2,q_o)={\bar\Omega\over q^2}.
\end{equation}
The $F$ factor for the ghost-gluon contribution to the ghost 
self-energy is
\begin{eqnarray}\label{ghgse}
F_{gh g}&=&\delta_{a b}N_ck^{\mu}q^{\nu}\left(g_{\mu \nu}-
(1-a){(k_{\mu}-q_{\mu})(k_{\nu}-q_{\nu})
\over (k-q)^2\pm 2i(k_o-q_o)\epsilon}\right)\cr
\nonumber\\
&\rightarrow &\delta_{a b}N_c{q^2 \over 2}.
\end{eqnarray}
The factor $q^2$ ensures the absence of pinch singularity and a 
well-defined perturbative result.

\subsection{ Scalar-Photon Self-Energy
 Contribution to the Scalar Propagator}
Although the massless scalar boson 
interacting with a photon is not part of massless QCD ,
it is treated using the same methods.

The $K$ factor is defined as
\begin{equation}\label{SbKbarOS}
K(q^2,q_o)={\bar\Omega\over q^2}.
\end{equation}
The $F$ factor for the massless scalar-photon contribution to the 
scalar self-energy,
\begin{eqnarray}\label{scphot}
& &F_{s\gamma}=(q+k)^{\mu}(q+k)^{\nu}\left(g_{\mu \nu}-
(1-a){(k-q)_{\mu}(k-q)_{\nu} \over
(k-q)^2\pm 2i(k_o-q_o)\epsilon}\right)\cr
\nonumber\\
& &=q^2\left(2-(1-a){q^2-k^2 \over 
(k-q)^2\pm 2i(k_o-q_o)\epsilon}\right)
\cr
\nonumber\\
& &\rightarrow 2q^2,
\end{eqnarray}
clearly exhibits the $q^2$ damping factor!

\section{ Conclusion}
Studying the out of equilibrium Schwinger-Dyson equation, we
have found that ill-defined pinch-like expressions appear 
exclusively in the Keldysh component (${\cal G}_K$) of the resummed 
propagator (\ref{sol1barG}), or in the single self-energy insertion 
approximation to it (\ref{pinch0p}).
This component does not vanish only in the expressions with
the Keldysh component (\ref{Omega}) ($\Omega $ or
$\bar\Omega $ for the single self-energy approximation)
of the self-energy matrix. This then requires that loop particles
be on mass shell. This is the crucial point to eliminate pinch 
singularities.

We have identified two basic mechanisms for the elimination of 
pinching: the threshold and the spin effects.

For a massive electron and a massless photon (or quark and gluon)
it is the threshold effect in the phase space integration
that produces, respectively, the critical $q^2-m^2$ or $q^2$
damping factors.

In the case of a massless quark, ghost, and gluon, this
mechanism fails, but the spinor/tensor structure of the self-energy
provides an extra $q^2$ damping factor.

We have found that, in QED, the pinching singularities appearing
in the single self-energy insertion approximation to the electron 
and the photon propagators are absent under very reasonable 
conditions: the distribution function should be finite, 
exceptionally the photon distribution is allowed to diverge as
$k_o^{-1}$ as $k_o\rightarrow 0$; the
derivative of the electron distribution should be finite; the 
total density of electrons should be finite.

For QCD, identical conditions are imposed on the distribution of 
massive quarks and the distribution of gluons; the distributions of 
massless quarks and ghosts (observe here that in the covariant 
gauge, the ghost distribution is not required to be identically zero) 
should be integrable functions;
they are limited  by the finiteness of the total density.

In the preceding sections we have shown that all pinch-like
expressions appearing in QED and QCD (with massless and massive 
quarks!)  at the single self-energy insertion level
do transform into well-defined expressions. Many other theories
behave in such a way.
However, there are important exceptions:
all theories in which lowest-order processes are
kinematically allowed do not acquire well-defined expressions at 
this level. These are electroweak interactions, processes 
involving Higgs  and two light particles, a $\rho $ meson and two
$\pi $ mesons, $Z$, $W$, and other heavy particles decaying into 
a pair of light particles, etc. The second important exception is 
massless $g^2\phi^3$ theory. This theory, in contrast to
massless QCD, contains no spin factors to provide 
(\ref{rpinchlike2}).
In these cases, one has to resort to the resummed Schwinger-Dyson
series.

The main result of the present paper is the cancellation of pinching
singularities at the single self-energy insertion level in QED- and
QCD-like theories. This, together with the
reported$^{\cite{lebellac,niegawacom}}$ cancellation of collinear
singularities, allows the extraction of useful physical 
information contained in the imaginary parts of the two-loop diagrams. 
 This is not the case with three-loop diagrams, because some of 
them contain double self-energy insertions. In this case, one again
has to resort to the sophistication of resummed propagators.

\vspace*{20mm}
{\large \bf Acknowledgment}
I acknowledge discussions with
R.~Baier and M.~Dirks. I also acknowledge financial support
from the EU under contract CI1$^*$-CT91-0893(HSMU). This work was
supported by the Ministry of Science and Technology of the
Republic of Croatia under Contract No. 00980102.
\newpage
\section{ Appendix}
We start$^{\cite{landsman}}$ by defining a heat-bath four-velocity
$U_{\mu}$, normalized to unity, and define the orthogonal projector
\begin{equation}\label{Delta}
\Delta_{\mu \nu}=g_{\mu \nu}-U_\mu U_\nu.
\end{equation}
We further define spacelike vectors in the heat-bath frame:
\begin{equation}\label{kappa}
\kappa_{\mu}=\Delta_{\mu \nu}q^\nu,~~~~\kappa_\mu \kappa^\mu
=\kappa^2=-\vec q^2.
\end{equation}

There are four independent symmetric tensors (we distinguish
retarded from advanced tensors by the usual
modification of the $i\epsilon $ prescription)
$A$, $B$, and $D$ (which are mutually orthogonal projectors), and $C$:
\begin{equation}\label{A}
A_{\mu \nu}(q)=
\Delta_{\mu \nu}-{\kappa_\mu \kappa_\nu \over \kappa^2},
\end{equation}
\begin{equation}\label{B}
B_{R~\mu \nu}(q)=U_\mu U_\nu +{\kappa_\mu \kappa_\nu
\over \kappa^2}-{q_\mu q_\nu \over (q^2+2iq_o\epsilon)},
\end{equation}
\begin{equation}\label{Cmunu}
C_{R~\mu \nu}(q)={(-\kappa^2)^{1/2} \over U.q}
\left ({(U.q)^2 \over \kappa^2}U_\mu U_\nu -{\kappa_\mu \kappa_\nu
 \over \kappa^2}+{q_o^2+\vec q^2 \over \vec q^2}
 {q_\mu q_\nu \over q^2+2iq_o\epsilon}\right ),
\end{equation}
\begin{equation}\label{Dmunu}
D_{R~\mu \nu}(q)={q_\mu q_\nu \over q^2+2iq_o\epsilon}.
\end{equation}

In addition to the known multiplication$^{\cite{landsman,weldont}}$
properties
\begin{equation}\label{AA}
A(q)A(q)=A(q),~~~~~~~~B_{R,A}(q)B_{R,A}(q)=B_{R,A}(q),
\end{equation}
\begin{equation}\label{CRCR}
C_{R,A}(q)C_{R,A}(q)=-(B_{R,A}(q)+D_{R,A}(q)),
~~D_{R,A}(q)D_{R,A}(q)=D_{R,A}(q),
\end{equation}
\begin{eqnarray}\label{AB}
& &A(q)B(q)=B(q)A(q)=0,~~A(q)C(q)=C(q)A(q)=0,
\cr
\nonumber\\
& &
A(q)D(q)=D(q)A(q)=0,~~B(q)D(q)=D(q)B(q)=0,
\end{eqnarray}
\begin{equation}\label{BC}
(B_{R,A}(q)C_{R,A}(q))_{\mu\nu}=(C_{R,A}(q)D_{R,A}(q))_{\mu\nu}
={\tilde q_\mu q_\nu\over q^2\pm 2iq_o\epsilon},
\end{equation}
\begin{equation}\label{CB}
(C_{R,A}(q)B_{R,A}(q))_{\mu\nu}=(D_{R,A}(q)C_{R,A}(q))_{\mu\nu}
={q_\mu \tilde q_\nu\over q^2\pm 2iq_o\epsilon},
\end{equation}
we need mixed products
\begin{equation}\label{BRBA}
B_{R,A}(q)B_{A,R}(q)={1 \over 2}(B_R(q)+B_A(q)),
\end{equation}
\begin{equation}\label{CRCA}
C_{R,A}(q)C_{A,R}(q)=-{1 \over 2}(B_R(q)+B_A(q)
+D_R(q)+D_A(q)),
\end{equation}
\begin{equation}\label{DRDA}
D_{R,A}(q)D_{A,R}(q)={1 \over 2}(D_R(q)+D_A(q)),
\end{equation}
\begin{equation}\label{BCRA}
(B_{R,A}(q)C_{A,R}(q))_{\mu\nu}=(C_{R,A}(q)D_{A,R}(q))_{\mu\nu}
={1\over 2}({\tilde q_\mu q_\nu\over q^2+2iq_o\epsilon}
+{\tilde q_\mu q_\nu\over q^2-2iq_o\epsilon}),
\end{equation}
\begin{equation}\label{CBRA}
(C_{R,A}(q)B_{A,R}(q))_{\mu\nu}=(D_{R,A}(q)C_{A,R}(q))_{\mu\nu}
={1\over 2}({q_\mu \tilde q_\nu\over q^2+2iq_o\epsilon}
+{q_\mu \tilde q_\nu\over q^2-2iq_o\epsilon}).
\end{equation}
By calculating the traces of the tensors $l^{\mu}l^{\nu}$,
$q^{\mu}l^{\nu}+l^{\mu}q^{\nu}$, and $q^{\mu}q^{\nu}$
with projectors, we find
\begin{equation}\label{llAB}
l^{\mu}l^{\nu}=m^2{q_o^2 \over 2\vec q^2}A^{\mu \nu}(q)
+{q^2 \over \vec q^2}
\left({4\vec l^2-q_o^2 \over 8}
A^{\mu \nu}(q)-l_o^2B^{\mu \nu}(q)\right)+O^{\mu\nu}(\vec k_T),
\end{equation}
\begin{equation}\label{qlC}
q^{\mu}l^{\nu}+l^{\mu}q^{\nu}=
-{q^2l_o\over |\vec q|}C^{\mu \nu}(q)+O^{\mu\nu}(\vec k_T),
\end{equation}
\begin{equation}\label{qqDB}
q^{\mu}q^{\nu}=q^2D^{\mu \nu}(q),
\end{equation}
\begin{equation}\label{gAB}
g^{\mu \nu}=(A^{\mu \nu}(q)+B^{\mu \nu}(q)+D^{\mu \nu}(q)).
\end{equation}
The tensor $O^{\mu\nu}(\vec k_T)$ is linear
in $\vec k$, and vanishes after integration over $\phi $.
One should observe that (\ref{llAB}) to (\ref{gAB}) are valid
for an arbitrary (but the same for $B$ and $D$) $R/A $ prescription,
so we do not indicate it.

Using the multiplication rules one obtains
\begin{eqnarray}\label{gfg}
& &(g_{\mu\rho}-(1-a)D_{R\mu\rho}(q))
(f_aA^{\rho \sigma}+f_bB^{\rho \sigma}
+f_cC^{\rho \sigma}+f_dD^{\rho \sigma}
(g_{\sigma\nu}-(1-a)D_{A\sigma\nu}(q))
\cr
\nonumber\\
& &=
{1\over 2}\biggl(f_aA_R^{\rho \sigma}+f_bB_R^{\rho \sigma}
-(1-a)f_cC_R^{\rho \sigma}+(1-a)^2f_dD_R^{\rho \sigma}\biggr)
\cr
\nonumber\\
& &+{1\over 2}\biggl(
f_aA_A^{\rho \sigma}+f_bB_A^{\rho \sigma}
-(1-a)f_cC_A^{\rho \sigma}+(1-a)^2f_dD_A^{\rho \sigma}\biggr).
\end{eqnarray}
It is important to observe that, owing to the properties of
the mixed products (\ref{BRBA}) to (\ref{CBRA}),
the $R/A$ assignment of $F$ does not influence
the final result!

\newpage
\end{document}